\documentclass[10pt,conference]{IEEEtran}
\IEEEoverridecommandlockouts
\usepackage{cite}
\usepackage{amsmath,amssymb,amsfonts}
\usepackage{algorithmic}
\usepackage{fancyhdr}
\usepackage{graphicx}
\usepackage{textcomp}
\usepackage[table]{xcolor}
\usepackage{pdfpages}
\usepackage{bm}
\def\BibTeX{{\rm B\kern-.05em{\sc i\kern-.025em b}\kern-.08em
    T\kern-.1667em\lower.7ex\hbox{E}\kern-.125emX}}
    
\fancypagestyle{specialfooter}{%
\fancyhf{}

\fancyfoot[R]{ \noindent\fbox{%
\parbox{\textwidth}{%
{\footnotesize \bf \tiny{© 2025 IEEE.  Personal use of this material is permitted.  Permission from IEEE must be obtained for all other uses, in any current or future media, including reprinting/republishing this material for advertising or promotional purposes, creating new collective works, for resale or redistribution to servers or lists, or reuse of any copyrighted component of this work in other works.}}
}
}}
}
    
\usepackage[margin=0.74in]{geometry}

\begin{document}


\twocolumn

\title{Grover's algorithm for image edge detection}

\author{\IEEEauthorblockN{Pulak Ranjan Giri}
\IEEEauthorblockA{
\textit{KDDI Research, Inc.}\\
Fujimino-shi, Saitama, Japan \\
pu-giri@kddi-research.jp}
\and
\IEEEauthorblockN{Kazuhiro Saito}
\IEEEauthorblockA{
\textit{KDDI Research, Inc.}\\
Fujimino-shi, Saitama, Japan \\
ku-saitou@kddi-research.jp}
}

\maketitle\thispagestyle{specialfooter}

\maketitle

\begin{abstract}
Grover's  algorithm  is one of the most significant  quantum algorithms,  enabling the   search for a marked element in an unsorted database  quadratically faster than  classical  search algorithms.   We utilize   Groves's   algorithm  to detect   the  edges  of  a digital  image  faster than any  classical edge detection methods.  The high success probability of our method in detecting  image edges,  compared to  most  existing quantum edge detection methods,  makes it a more suitable  candidate for use  in the presence of noise   on  current  quantum devices.  We implemented    our  model  for  the  $2\times 2$ blocks of  the image  using   a  two-qubit  Qiskit  circuit   with  the   QASM simulator and  the noisy IBM Sydney(fake)  device,   making it suitable for  most of the  current NISQ devices.

\end{abstract}


\section{Introduction} \label{in}
Edge detection  \cite{sobel} of a digital image\textemdash an  important  task  in  image processing\textemdash  is the process of locating the  gradients in  pixel intensities within  an image.  It has various applications,  including 
medical imaging, forensic science,  material science,   and traffic surveillance.  Classical edge detection algorithms   \cite{pre,kir,can,niya} often struggle   to process  large   images as they require an  exponentially  time-consuming method of  examining  each pixel for the  image edge  detection,  which is  detrimental to applications such as    self-driving vehicles \cite{thai},  where real-time  edge detection is    necessary for a safe driving experience. 

In this regard,  fast detection  of image edges  by  quantum algorithms  \cite{ni}  could be an alternative to the slow classical  methods.  
Grover's  algorithm  \cite{grover1}  is one of  the  quantum algorithms that  can search for marked elements  in  an unsorted database  quadratically faster than  classical search algorithms.

Existing quantum edge detection methods  such as  QSobel \cite{zhang} and Hadamard edge detection (HED) \cite{wei}, though faster than the classical methods,  face challenges due to the low gradient intensities at the image edges, making it difficult to identify edges  specifically in the  presence of  noise.    In this regard,  our recent method  \cite{giripla}  based on quantum walk search can easily  identify edges  with high success probability.

In this article,  we propose an alternative   quantum edge detection method  utilizing   Grover's  algorithm  that requires less quantum resources than our previous method \cite{giripla}.   This   method,  similar to our previous one,  searches for the edge pixels as marked elements  with high success probability.

This  article  is organized     as follows:   Edge detection utilizing Grover's algorithm   is  presented  in Section    \ref{eg}, which includes   Grover's algorithm  in Subsection \ref{grovalg}, Qiskit implementation  in Subsection \ref{qiskit} and experimental results  in Subsection \ref{exp}.    Finally,   we conclude in Section   \ref{con} with a discussion. 
\section{Edge detection by Grover's algorithm} \label{eg}
 In this section we discuss how we can utilize  Grover's algorithm to  search for  the edge pixels  of  a   $L_1 \times L_2$ digital image   with  pixel intensities $I(x,y)$ at  $(x,y)$  coordinates.  
Let us assume that the  $M$ edge  pixels of the  image  belong  to  the  set  $\mathcal{T}_M$ of marked elements,   which can be obtained  using a  one-dimensional filter mask, as discussed in detail in  \cite{giripla} and refs. therein.    The horizontal and vertical gradients $I_{h\pm}(x,y) =  I(x,y) - I(x\pm1, y) $  and $I_{v\pm}(x,y) =  I(x,y) - I(x, y\pm1) $ respectively  are used to obtain 
\begin{eqnarray}
I_{max}(x,y) =  \mbox{max} \left( I_{h+}, I_{h-}, I_{v+}, I_{v-} \right)\,,
\label{thcon1}
\end{eqnarray}
which  is useful to determine the existence of edges if it satisfies the threshold condition  $I_{max}(x,y)  \geq a_{th}$, 
where $a_{th}$ is the pre-determined threshold value, that depends on the image.  Based on the result in   (\ref{thcon1}),  the marking of edges are done by the oracle of  Grover's  iteration  operator, which are subsequently  amplified and finally measured.   Bellow,  we  discuss  Grover's algorithm  for  the purpose  of edge detection of a digital image.
\subsection{Grover's algorithm} \label{grovalg}
Grover's algorithm can search for $M$ edge pixels  of the digital image from  the  set of  $N = L_1 L_2$  pixels,  present in the image, quadratically faster than any classical search algorithms.  
The algorithm starts with the initial state, $|\psi_{in}\rangle$,   which is the equal superposition of all the $N$ basis  states as
\begin{eqnarray}
 |\psi_{in}\rangle =   \frac{1}{\sqrt{N}} 
\sum_{i=0}^{N-1} |i \rangle\,,
 \label{in}
\end{eqnarray}
where  $N$  basis states  $|i \rangle = | xL_2+y \rangle$ are obtained by concatenating all the rows of the    $L_1 \times L_2$  lattice. 
The Grover's  iteration operator (evolution operator) is given by 
\begin{eqnarray}
\mathcal{G} =   \mathcal{I}_{\psi} \mathcal{I}_{t}  \,,
 \label{in}
\end{eqnarray}
where   the diffusion operator   $\mathcal{I}_{\psi} = 2 |\psi_{in}\rangle \langle \psi_{in}| - \mathbb{I}$ is  the reflection term   about the double average of all the probability amplitudes of the state vector   and  the oracle  operator  $\mathcal{I}_{t} = \mathbb{I} - 2  \sum_{i} | t_i \rangle \langle t_i |$  reflects about the target elements $t_i$.  
The final state, $|\psi_{f}\rangle$,  is obtained after  $T= \frac{\pi}{4}\sqrt{N/M}$ applications  of the Grover iteration operator to the initial state, $|\psi_{in}\rangle$,  as 
\begin{eqnarray}
 |\psi_{f}\rangle =   \mathcal{G}^T |\psi_{in}\rangle\,.
 \label{fn}
\end{eqnarray}
The   measurement result  $p_i =  |\langle t_i |\psi_{f}\rangle|^2$ gives the  edges of the digital image with high success probability. 
\begin{figure}
 \includegraphics[width=0.4\textwidth]{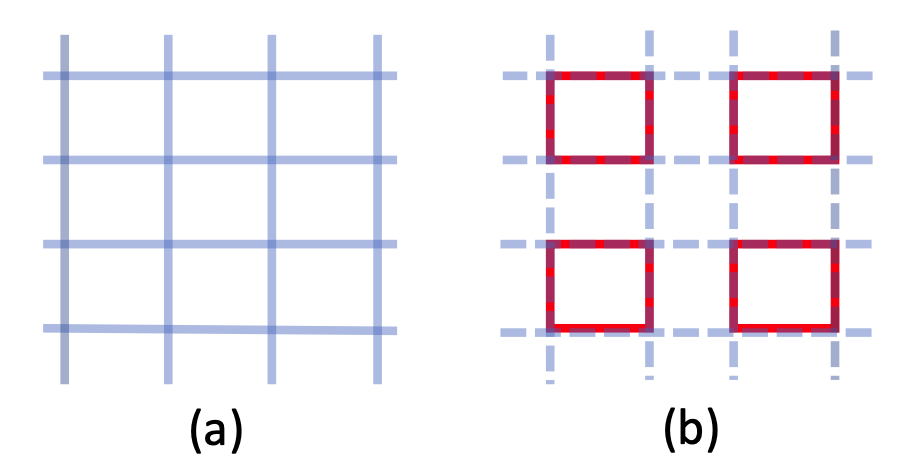}
          
       \caption{\bf{(a) A  $4\times 4$ digital image represented as a two-dimensional lattice of size $4\times 4$, (b) four $2\times 2$ blocks of the  $4\times 4$ two-dimensional lattice.}}
 \label{f1}
\end{figure}

\begin{figure}[h!]
  \centering
     \includegraphics[width=0.35\textwidth]{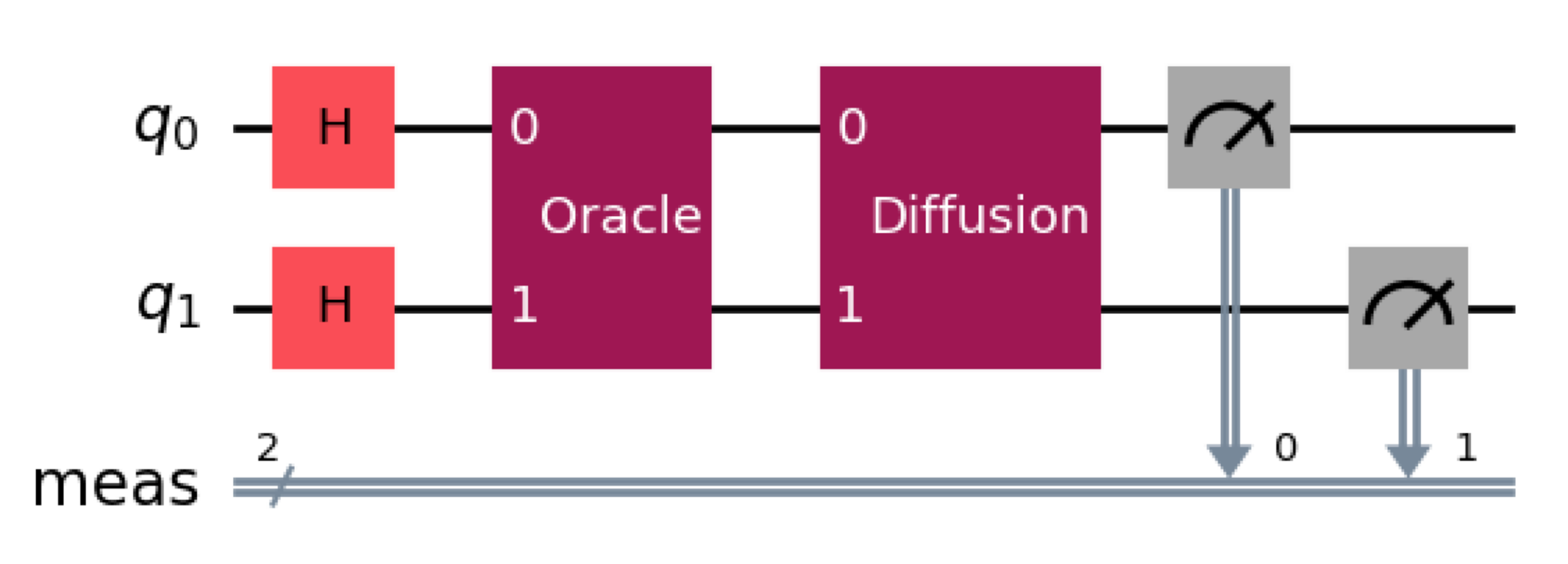}
          
       \caption{Qiskit circuit for  Grover's search algorithm with $N=4$ elements  for the quantum edge detection.}
 \label{f2}
\end{figure}

\begin{figure}
  \centering
     \includegraphics[width=0.35\textwidth]{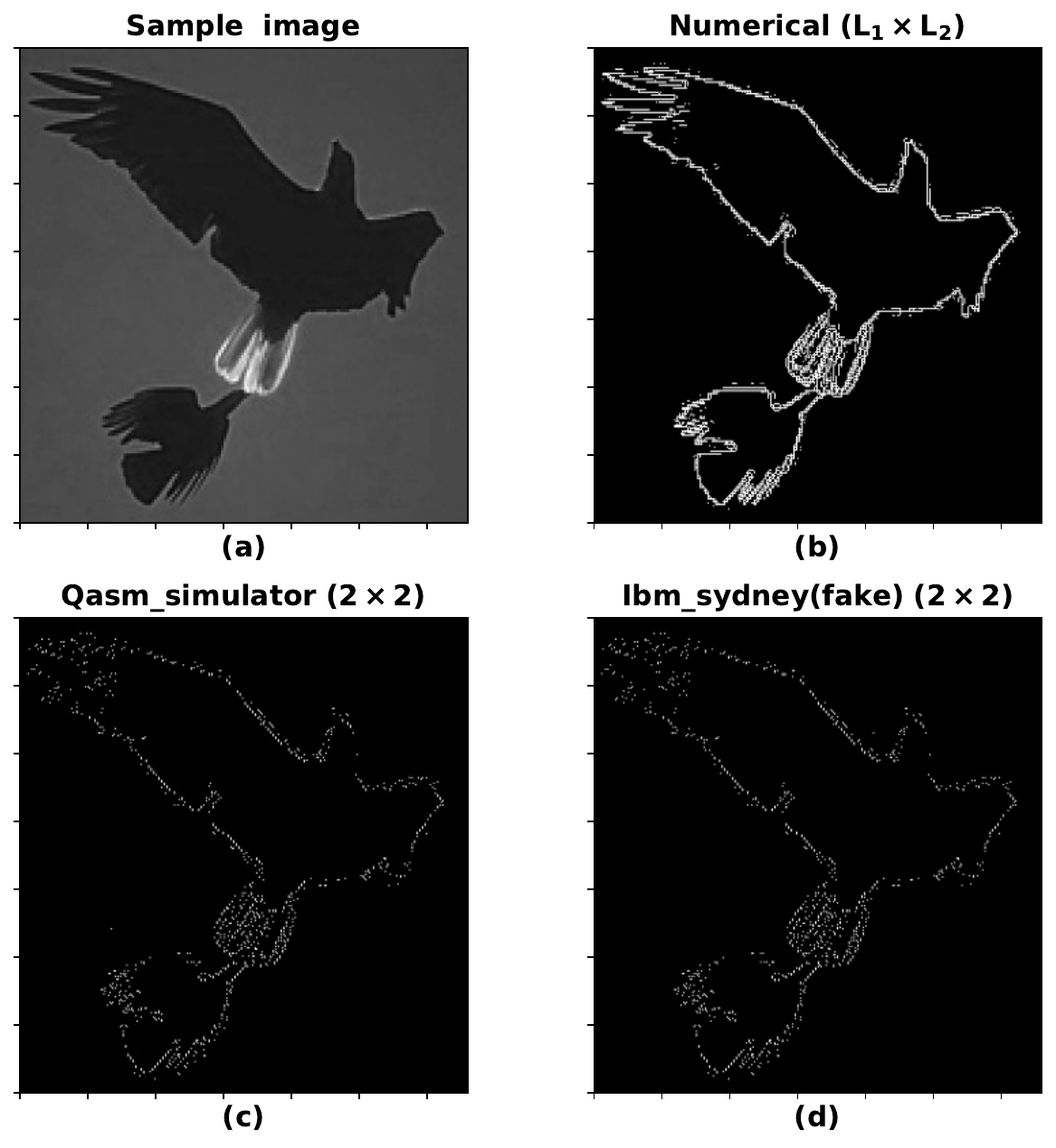}
          
       \caption{\bf{(a) A  $\bf  L_1 \times L_2= 330 \times 350$ image  taken from the BSDS500 database,  (b) edge detected by numerical analysis  of  Grover's search on the full image,  edge detected by  Grover's search  on $2\times 2$ blocks of the image with (c) $qasm\_simulator$ and with (d) $ibm\_sydney(fake)$ device with noise.}}
 \label{f4}
\end{figure}
\subsection{Qiskit Implementation} \label{qiskit}
The implementation of  Grover's algorithm for the edge detection of an entire  image with a   reasonable size  is not feasible  in the  present NISQ-era quantum computers.   
Therefore, we adopt a strategy to divide the image   in Fig. \ref{f1}(a)  into $2\times 2$ red blocks as  shown in Fig. \ref{f1}(b) to  perform  Grover's search on each of the   $2\times 2$ blocks, which is scalable without additional  overhead on the quantum computer.   
Grover's search then  runs  with  $N =4$  elements  for  $T =1$ time  with shots $= 1024$  as presented by the Qiskit circuit in Fig. \ref{f2}. 
Measurement of the success probabilities  from all the blocks  are mapped to the  $L_1 \times L_2$ lattice, which  represents   the edges of the original image.  Dividing the image into  bigger blocks, which is  also possible to run in IBM quantum devices with more qubits, will provide better results.  
\subsection{Experimental results} \label{exp}
The result of our experiment is presented in Fig. \ref{f4},  with the sample image  in Fig. \ref{f4}(a)  along with  three  corresponding  image  edges.    The image edge  in  Fig. \ref{f4}(b) is obtained from the numerical analysis of  Grover's search considering  the  entire image as a single  search problem.  The image edge  in  Fig. \ref{f4}(c) is obtained  using  $qasm\_simulator$  with   Grover's search  on $2\times 2$ blocks of the image, and  in  Fig.  \ref{f4}(d), the  same method is performed   using  $ibm\_sydney(fake)$ device. 
 
\section{Conclusions} \label{con}
In this article,  we  propose  an  edge detection method  utilizing Grover's search algorithm,  which requires  fewer  qubit resources than our previous model \cite{giripla} based on quantum walk search. For example, for the execution of a $2\times 2$  pixel block of an image this method requires two qubits, while our previous method requires four qubits.  Existing  quantum methods such as QSobel \cite{zhang} and HED  \cite{wei}, which  measure  only pixel gradients  for the edge detection often struggle with  low success probability, especially   in noisy environments.    The present  approach using  Grover's  search   provides a  higher  success probability,  making it more robust against noise.     A  small Qiskit  implementation   presented in this article  can  be run  on the noisy  IBM quantum computer.  We hope that the  use of fixed-point  Grover's search \cite{ch}, which is more suitable for  unknown number of marked elements, and search for the whole image at a time instead of small blocks  in quantum devices   can  further enhance the performance of our  model. 


\end{document}